\documentstyle[preprint,pra,aps]{revtex}
\begin{document}
\draft
\title{\bf  Precise calculation of parity nonconservation
in cesium and test of the standard model}
\author{V.A. Dzuba, V.V. Flambaum, and J.S.M. Ginges}
\address{School of Physics, University of New South Wales,
Sydney 2052, Australia}
\date{\today}
\maketitle

\tightenlines

\begin{abstract}
We have calculated the $6s-7s$ parity nonconserving (PNC) E1 transition
amplitude, $E_{PNC}$, in cesium.
We have used an improved all-order technique in the calculation
of the correlations and have included all significant contributions
to $E_{PNC}$.
Our final value
$E_{PNC}=0.904\Big( 1\pm 0.5\%\Big) \times 10^{-11}iea_{B}(-Q_{W}/N)$
has half the uncertainty claimed in old calculations
used for the interpretation of Cs PNC experiments.
The resulting nuclear weak charge $Q_{W}$ for Cs deviates by
about $2\sigma$ from the value predicted by the standard model.

\end{abstract}
\vspace{1cm}

\pacs{PACS: 11.30.Er, 12.15.Ji, 32.80.Ys, 31.30.Jv}

\section{Introduction}

There is an ongoing discussion on whether measurements of parity
nonconservation (PNC) in atoms can be used to study new physics beyond the
standard model (for a history of PNC in atoms, see, e.g.,
the book \cite{khriplovich} or review \cite{bouchiat}).
Such a possibility relies on the accuracy of
the measurements and the theoretical analysis.
The best accuracy so far has been achieved for the cesium atom.
In 1997, the Boulder group measured the PNC amplitude in cesium
to an accuracy of 0.35\% \cite{wood97}. The best calculations available
at that time were published by the Novosibirsk group in 1989 \cite{dzuba89}
and the Notre Dame group in 1990 \cite{blundell90}.
Both works claimed an accuracy of 1\%.
This claim of 1\% accuracy was based in part on the comparison
of the experimental and theoretical values relevant to the PNC amplitude
(energies, electromagnetic E1 transition amplitudes, hyperfine structure).
Later, new, more accurate measurements (see, e.g. \cite{rafac,bennett99})
led to better agreement between theory and experiment for the
E1 transition amplitudes.
This allowed Bennett and Wieman to suggest that the actual
accuracy of the PNC calculations in cesium is 0.4\% \cite{wieman99}.
With this accuracy, the value of the nuclear weak charge $Q_{W}$
of the cesium atom which follows from the measurements \cite{wood97} and
calculations \cite{dzuba89,blundell90} deviates from the standard model
value by $2.5\sigma$ \cite{wieman99}.
The implications of this deviation for physics beyond the standard model
have been examined in several works \cite{RC,JR,EL}.

This result generated many recent works revisiting calculations of
PNC in cesium.
Derevianko \cite{derevianko00} demonstrated that the contribution
of the Breit interaction to the PNC amplitude $E_{PNC}$ (-0.6\%) is
substantially larger than previous estimates and reduces the deviation
from the standard model. The Breit contribution was neglected in \cite{dzuba89}
and underestimated in \cite{blundell90}. The result of Derevianko was
later confirmed in independent calculations \cite{harabati,kozlov01}.
A new many-body calculation of  $E_{PNC}$ was performed by
Kozlov {\it et al.} \cite{kozlov01}.
It was calculated in second-order in the residual interaction
with averaged screening factors.
The result is in excellent agreement with similar calculations by
Blundell {\it et al.} \cite{blundell90} and differs by less
than 0.5\% from the more complete (``all-orders'') calculations
\cite{dzuba89,blundell90}.

Radiative corrections to the weak charge
of order $\alpha$ were calculated a long time ago
in \cite{marciano} (see also \cite{lynnbed}).
However, there are important ``strong field'' radiative corrections
of order $Z\alpha ^{2}$ and $Z^2\alpha ^{3} \ln^2(\lambda/R_n)$
which have recently been considered in
\cite{milstein} (note that the latter correction is larger).
Here $\lambda$ is the electron Compton wavelength, and
$R$ is the nuclear radius.
The main contribution to $E_{PNC}$ was found to be about $0.4\%$.
This contribution originates from the radiative corrections to the
weak matrix element due to the Uehling potential
(recently this contribution was calculated numerically in \cite{johnson01}).

Among other corrections considered
one should mention a correction for the neutron distribution
\cite{derevianko} which is quite small: -0.2\% of $E_{PNC}$.

Note that Breit, Uehling, and neutron skin corrections are
smaller than the uncertainty of 1\% claimed for the calculated $E_{PNC}$
amplitude in works \cite{dzuba89,blundell90}. This uncertainty was
mostly associated with the accuracy of the calculations of correlation
corrections. Therefore, there is a very important question as to
whether calculations of correlations were, or can be, performed to
better than 1\% accuracy.
The analysis of the accuracy  involves a comparison of
calculated and measured atomic quantities
{\em and} some additional  ``internal'' tests, such as, e.g.,
checking the stability of the results against variation of certain
parameters.
Comparison with experiment alone is an important
but not sufficient part of the analysis. This is especially true when
better accuracy is claimed for calculations performed many years
ago and all the details can hardly be reconstructed. In our view,
the only reliable way to improve the accuracy is to do the calculations
again, trying to improve the method and numerical procedures at every
stage, and repeat the analysis of the accuracy.
That is what we do in the present paper.

We have made several important improvements to the
method developed in 1989 \cite{dzuba89}. First, we calculated a new series
of higher-order correlation diagrams which account for the effect of
screening of the Coulomb interaction in the exchange correlation diagram.
In our earlier work \cite{dzuba89} the screening effect was calculated
for the direct correlation diagram only, while screening factors
were used for the exchange diagram.
The values of these factors were found by looking at the effect of
screening in the direct diagram; however, the effect
of screening in the exchange diagram may differ.
In our present work we calculate the effect of screening for both diagrams.

Second, we have made significant improvements to the numerical procedures
at every stage of the calculations. This involves, for instance,
using a more dense coordinate grid, including more terms
into the summation over virtual states,
and restoring the lower Dirac component of the wave function everywhere.
We have also used a more accurate nuclear charge distribution.

Our result for the PNC amplitude in cesium
(without Breit, radiative, and neutron distribution corrections)
$E_{PNC}=0.908\Big( 1\pm 0.5\% \Big) \times 10^{-11}iea_{B}(-Q_{W}/N)$
coincides exactly with the old result
\cite{dzuba89} but has a much smaller uncertainty.
($N$ is the neutron number.)
To avoid misunderstanding, we should note
that particular contributions to this value are slightly different
in the old and the new calculations (for example, the Hartree-Fock
value of the amplitude has changed by 0.3\%).
Therefore, we are indeed talking about a new result.

We have also calculated the contributions of the
Breit interaction (-0.61\%) and the Uehling potential (0.40\%)
to $E_{PNC}$. There is very good agreement between different calculations
for these values and they can be considered well-established.
The resulting value for the PNC amplitude is
$E_{PNC}=0.906\Big( 1\pm 0.5\% \Big)\times 10^{-11}iea_{B}(-Q_{W}/N)$.
The neutron-distribution correction,
estimated as $(-0.2 \pm 0.1)\%$, reduces our final value to
$E_{PNC}=0.904\Big( 1\pm 0.5\% \Big) \times 10^{-11}iea_{B}(-Q_{W}/N)$.

\section{Method of calculation}
\label{sec:method}

The method we have used here is very similar to that used in our 1989
work \cite{dzuba89}.
As a zero approximation we use the relativistic Hartree-Fock
method. The perturbation in our approach is the residual
interaction - the difference between the exact and Hartree-Fock
Hamiltonians.
For summing of diagrams we use a combination of the ``correlation potential
method'' \cite{dzuba87}, which is a way to treat correlations by introducing
a single-electron operator (correlation potential) $\hat \Sigma$,
and ``perturbation theory in the screened electron residual interaction''
\cite{dzuba89energy,dzuba89e1hfs}, which is used to calculate $\hat \Sigma$.
The method is an all-order technique, in terms of treating correlations,
and is not a version of the popular coupled-cluster approach.
The dominating sequences of higher-order correlation diagrams
correspond to real physical phenomena like collective screening of the
Coulomb interaction between electrons and the hole-particle interaction.
They are included in all orders in our technique.
An important strong point of the method is that its complexity does not go
beyond the calculation of the correlation potential $\hat \Sigma$.
When $\hat \Sigma$ is ready, the calculation of energies and
matrix elements is relatively simple.
$\hat{\Sigma}$ is used to calculate single-electron Brueckner orbitals.
The calculation of matrix elements with Brueckner orbitals
 already includes most of the correlations.
There are additional ``non-Brueckner'' contributions to matrix elements such
as structural radiation and normalization of many-electron states. These
contributions are small for $s$ and $p$ states and in most cases they
can be  expressed in terms of derivatives of $\hat \Sigma$.
This makes the calculation of hyperfine structure, transition amplitudes,
etc. similar to the calculation of energies.

The method works very well for alkaline atoms. The calculated
energy levels and transition amplitudes deviate by a fraction
of a percent from experiment.
The hyperfine structure (hfs) of $s$ and $p$ states has an
accuracy of better than 1\%.

\subsection{Correlation potential method}

The Cs atom has one electron above closed shells. Therefore, it is natural
to start the calculations from the relativistic Hartree-Fock (RHF) method
in the $\hat{V}^{N-1}$ approximation (one electron in the field of the
$N-1$ core electrons).
The single-electron RHF Hamiltonian is
\begin{equation}
\label{eq:RHF}
\hat{H}_{0}=c\mbox{\boldmath$\alpha$}\cdot \hat{{\bf p}}+(\beta -1)c^{2}-
Z\alpha/r+\hat{V}^{N-1} \ ,
\end{equation}
$\mbox{\boldmath$\alpha$}$ and $\beta$ are Dirac matrices and
$\hat{{\bf p}}$ is the electron momentum.
For cesium the accuracy of the RHF energies is of the order of $10\%$.

To improve the accuracy one needs to include correlations.
In order to do so we introduce a ``correlation potential'' $\hat \Sigma$
\cite{dzuba87}.  $\hat \Sigma$ is a single-electron energy-dependent
non-local operator defined in such a way that its average value over
a single-electron state $a$ is the correlation correction to the energy
of this state,
\begin{equation}
\label{eq:S}
	\Delta \epsilon_a = \langle a | \hat \Sigma(\epsilon_a) | a \rangle.
\end{equation}
The calculation of $\hat \Sigma$ will be described in the next section.
$\hat \Sigma$ is added to the Hartree-Fock Hamiltonian to calculate
single-electron states of the external electron,
\begin{equation}
\label{eq:Br}
	(\hat H_0 + \hat \Sigma - \epsilon) \psi^{\rm Br} = 0.
\end{equation}
There are two major benefits from inclusion of $\hat \Sigma$ in the equations.
First, the iteration of $\hat \Sigma$ is an important higher-order effect
which gives a significant contribution to the energy and wave function.
Second, by solving Eq.~(\ref{eq:Br}) we obtain single-electron orbitals
which are called Brueckner orbitals which already include correlations.
These orbitals can then be used to calculate matrix elements for
hfs, PNC, etc.

When appropriate higher-order diagrams are included into the calculation
of $\hat \Sigma$, as will be described in the next section, by solving
Eq.~(\ref{eq:Br}) we obtain energies of $s$ and $p$ states of alkaline
atoms with an accuracy of about 0.1\%. This is a radical improvement
over the 10\% accuracy of the Hartree-Fock approximation.

To accurately calculate the interaction of external fields with atomic
electrons one needs also to take into account the core polarization effect.
We do this using the time-dependent Hartree-Fock (TDHF) method
(which is equivalent to the random-phase approximation with exchange).
In this method every single-electron orbital has the form
\[ \psi_n + \delta \psi_n \ , \] where $\psi_n$ is the unperturbed
Hartree-Fock orbital (or Brueckner one for states above the core)
and $\delta \psi_n$ is a correction to the orbital caused by
the external field.
To find $\delta \psi_n$ one needs to solve the equations
\begin{equation}
\label{eq:RPA}
	(\hat H_0 - \epsilon)\delta \psi_n = - (\delta \epsilon +
	\hat h + \delta \hat V_h) \psi_n.
\end{equation}
Here $\hat h$ is the operator of the external field, $\delta \hat V_h$
is the correction to the Hartree-Fock potential due to modification
of the core states in the external field, and
$\delta \epsilon$ is the correction to the energy.

Equations (\ref{eq:RPA}) are first solved self-consistently for core
states to obtain  $\delta \hat V_h$. Then Brueckner orbitals are used
to calculate matrix elements between states of the external electron
\begin{equation}
\label{eq:me}
	M_{ab} = \langle \psi^{\rm Br}_b |\hat{h} +\delta \hat{V}_{h}|
	\psi^{\rm Br}_a \rangle.
\end{equation}
This expression includes most of the correlations and already gives hfs
and transition amplitudes in alkaline atoms to an accuracy close to 1\%.
There are also contributions arising from the normalization of states and
structural radiation \cite{dzuba87}. When these contributions are included
the accuracy improves.

\subsection{Calculation of the correlation potential}

We use the Feynman diagram technique to calculate the correlation potential
$\hat \Sigma$. The main reason for this is that this technique is more
convenient for inclusion of higher-order diagrams. The drawback of it
is the necessity to integrate over frequencies numerically.

\begin{figure}[b]
\centerline{\psfig{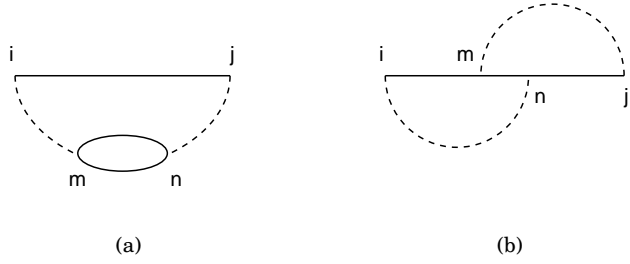}}
\caption{Second-order (a) direct and (b) exchange correlation diagrams
in the Feynman diagram technique.
Dashed line is the Coulomb interaction.
Loop is the polarization of the atomic core.}
\label{fig:Sigma2}
\end{figure}

In the lowest (second) order in the residual Coulomb interaction,
$\hat \Sigma$ is given by the direct and exchange diagrams in
Fig.~\ref{fig:Sigma2}.
The corresponding mathematical expressions are
\begin{eqnarray}
	\hat \Sigma^{(2)}_{\rm d} (\epsilon, r_i, r_j) &=&
	\sum_{nm} \int \frac{d\omega}{2\pi} G_{ij}(\epsilon+\omega)
	\hat Q_{im}\hat \Pi_{mn}(\omega)\hat Q_{nj} \label{eq:sigma2d},\\
	\hat \Sigma^{(2)}_{\rm ex} (\epsilon, r_i, r_j) &=&
	\sum_{mn} \int \int \frac{d\omega_1}{2\pi}\frac{d\omega_2}{2\pi}
	\hat Q_{in}G_{im}(\epsilon+\omega_1)G_{mn}(\epsilon+\omega_1+\omega_2)
	\hat Q_{mj}G_{nj}(\epsilon+\omega_2) \label{eq:sigma2ex},
\end{eqnarray}
where summation over $m$ and $n$ is a numerical implementation of the
integration over $r_m$ and $r_n$, $\hat Q$ is the Coulomb interaction,
$G$ is the Hartree-Fock Green function, $\hat \Pi$ is the
polarization operator:
\begin{eqnarray}
	&\hat Q_{ij} = \frac{e^2}{|r_i - r_j|}, \label{eq:q}\\
	&G_{ij}(\epsilon) =
	\sum_{\gamma} \frac{|\gamma\rangle_i\langle\gamma|_j}
	{\epsilon-\epsilon_{\gamma}+i\delta} +
	\sum_{n} \frac{|n\rangle_i\langle n|_j}
	{\epsilon-\epsilon_{n}-i\delta}, \ \ \delta \rightarrow 0,
	\label{eq:gf}\\
	&\hat \Pi_{ij}(\omega)=\int \frac{d\alpha}{2\pi}G_{ij}(\omega+\alpha)
	G_{ij}(\alpha) =\sum_n\psi_n^{\dagger}\big(
	G_{ij}(\epsilon_n+\omega)+G_{ij}(\epsilon_n-\omega)\big)\psi_n.
\label{eq:po}
\end{eqnarray}
Here $|n \rangle$ are core RHF states and $|\gamma \rangle$ are RHF excited
states;
$\delta$ shows the pole passing rule for integration over frequencies.
(Let us remind the reader that we use the Feynman diagram
technique to treat a basically non-relativistic correlation problem, e.g.,
our polarization operator (\ref{eq:po}) represents the polarization
of the atomic core, not the vacuum.)

The integration over frequencies in expressions (\ref{eq:sigma2d}) and
(\ref{eq:sigma2ex})
can easily be performed analytically, reducing the diagrams in
Fig.~\ref{fig:Sigma2}
to the four usual Goldstone diagrams presented, e.g.,
in work \cite{dzuba87}.
When higher-orders are included in $\hat \Sigma$,
only the integration for the polarization operator (\ref{eq:po}) can still be
done analytically. The other integrals have to be calculated numerically.

As was demonstrated in our earlier work \cite{dzuba89energy},
the most important higher-order correlation contributions to
$\hat \Sigma$ come from screening
of the Coulomb interaction between atomic electrons by other electrons
and from the hole-particle interaction in the polarization operator (another
important higher-order effect is the iteration of $\hat \Sigma$, which is
included separately, by inserting $\hat \Sigma$ into equations for
single-electron orbitals; see previous section).

\begin{figure}[b]
\centerline{\psfig{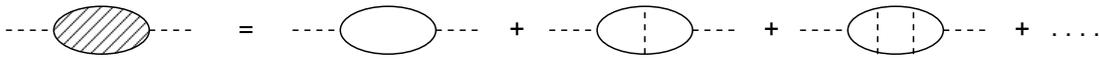}}
\caption{Hole-particle interaction in the polarization operator.}
\label{fig:h-p}
\end{figure}

The hole-particle interaction in the polarization operator $\hat \Pi$
(Fig.~\ref{fig:h-p}) is included by replacing the $\hat{V}^{N-1}$
Hartree-Fock potential
in the calculation of $\hat \Pi$ by the $\hat{V}^{N-2}$ potential. Since
the calculation involves iterations of the RHF equation, the hole-particle
interaction is included in all orders.

\begin{figure}[b]
\centerline{\psfig{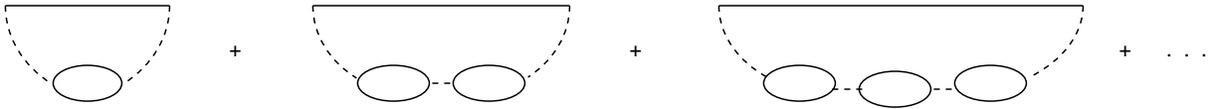}}
\caption{Screening of the Coulomb interaction
in the direct correlation diagram.}
\label{fig:Sigmad}
\end{figure}

The screening of the Coulomb interaction is a more complicated effect.
The corresponding higher-order diagrams can be obtained from the second-order
diagrams (Fig.~\ref{fig:Sigma2}) by the insertion of a series of hole-particle
loops into every Coulomb line. Therefore, instead of,
e.g., one direct diagram (Fig.~\ref{fig:Sigma2}) we have an infinite chain of
higher-order diagrams (Fig.~\ref{fig:Sigmad}). One can see that an internal
part of this chain (before the integration over $\omega$) forms a matrix
geometric progression
\[
	\hat \Pi + \hat \Pi \hat Q \hat \Pi +
	\hat \Pi \hat Q \hat \Pi \hat Q \hat \Pi + \dots
\]
The sum of this progression is
\begin{equation}
	\hat \pi(\omega)=
\hat \Pi(\omega)\big[ 1-\hat Q \hat \Pi(\omega)\big]^{-1},
\label{eq:screenedpi}
\end{equation}
and can be called the ``screened polarization operator''. Screening
of the Coulomb interaction in the direct correlation diagram is included by
replacing the unscreened polarization operator $\hat \Pi$ in
Eq.~(\ref{eq:sigma2d}) by the screened polarization operator $\hat \pi$:
\begin{eqnarray}
\label{eq:sigmad}
	\hat \Sigma_{\rm d} (\epsilon, r_i, r_j) &=&
	\sum_{nm} \int \frac{d\omega}{2\pi} G_{ij}(\epsilon+\omega)
	\hat Q_{im}\hat{\pi}_{mn}(\omega)\hat Q_{nj}.
\end{eqnarray}

\begin{figure}[b]
\centerline{\psfig{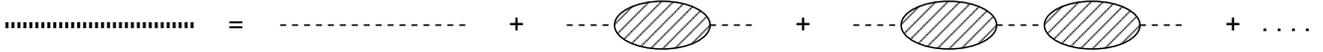}}
\caption{The screened Coulomb interaction
(with hole-particle interaction).}
\label{fig:screeninghp}
\end{figure}

It is convenient also to introduce an operator of the screened
Coulomb interaction
\begin{equation}
	\widetilde Q(\omega)=\hat Q (1-\hat Q \hat{\pi}(\omega))^{-1}.
\label{eq:screenedQ}
\end{equation}
This expression can be obtained in a similar way to the screened
polarization  operator $\hat \pi$ by summing a matrix geometric progression
corresponding to the chain of diagrams in Fig.~\ref{fig:screeninghp}.
An expression for the direct correlation diagram can be re-written in
terms of $\widetilde Q$:
\begin{eqnarray}
\label{eq:sigmade}
	\hat \Sigma_{\rm d} (\epsilon, r_i, r_j) &=&
	\sum_{nm} \int \frac{d\omega}{2\pi} G_{ij}(\epsilon+\omega)
	\hat Q_{im}\hat{\Pi}_{mn}(\omega)\widetilde Q_{nj}(\omega).
\end{eqnarray}
The screened Coulomb interaction operator $\widetilde Q$ can also
be used to include screening into the exchange correlation diagram.
This is done by substituting $\hat Q \rightarrow \widetilde Q(\omega)$ in
equation~(\ref{eq:sigma2ex})
\begin{eqnarray}
\label{eq:sigmaex}
	&&\hat \Sigma_{\rm ex} (\epsilon, r_i, r_j) = \nonumber \\
	&&\sum_{mn} \int \int \frac{d\omega_1}{2\pi}\frac{d\omega_2}{2\pi}
	\widetilde Q_{in}(\omega_1)G_{im}(\epsilon+\omega_1)
	G_{mn}(\epsilon+\omega_1+\omega_2)
	\widetilde Q_{mj}(\omega_2)G_{nj}(\epsilon+\omega_2).
\end{eqnarray}

\begin{figure}[b]
\centerline{\psfig{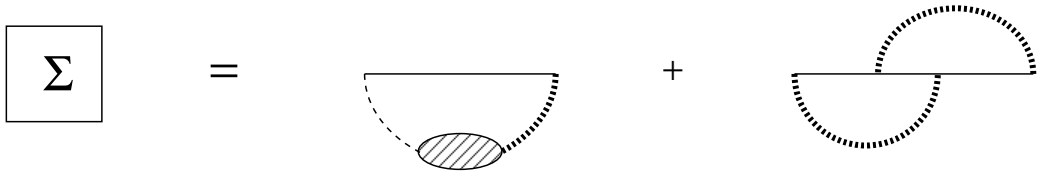}}
\caption{The electron self-energy operator with screening and hole-particle
interaction included.}
\label{fig:Sigmahpsc}
\end{figure}

The second-order correlation potential with the
hole-particle interaction and screening of the Coulomb interaction
included in all orders is depicted in Fig.~\ref{fig:Sigmahpsc}.

In our 1989 work \cite{dzuba89} we used an approximate expression for the
screened Coulomb interaction in the exchange correlation diagram
\begin{equation}
\label{eq:fk}
	\widetilde Q_k \approx f_k \hat Q_k,
\end{equation}
where $k$ is the multipolarity of the Coulomb interaction and $f_k$ is the
screening factor. In this expression the dependence of the screening on
frequency is neglected. The values of the screening factors were obtained by
calculating the direct diagram with and without screening.
The use of average screening factors which do not depend on frequency
significantly simplified the calculation of the exchange diagram. Like
in pure second-order, the Goldstone diagram technique and direct summation
over a complete set of single-electron states were used. No integration
over frequencies was needed.

In the present work we do the full-scale calculation of the exchange diagram
using Eq.~(\ref{eq:sigmaex}). This makes the method totally
{\it ab initio} since no screening factors are used. The drawback of
the method is the need to do double integration over frequencies numerically
(note that there is only a single numerical integration over frequencies
in the direct diagram).

\subsection{PNC amplitude}

To calculate the PNC $6s-7s$ amplitude in Cs one needs to include
two external fields acting on the atomic electrons:
the weak field of the nucleus and the electric dipole
(E1) field of the photon.
The nuclear spin-independent weak interaction of an electron with
the nucleus is
\begin{equation}
\label{eq:h_w}
\hat{H}_{W}=\frac{G_{F}}{2\sqrt{2}}\rho (r)Q_{W}\gamma _{5}
\end{equation}
where $G_{F}$ is the Fermi constant, $Q_{W}$ is the weak charge of the
nucleus, $\gamma _{5}$ is a Dirac matrix, and $\rho (r)$ is the nuclear
density. The E1 Hamiltonian is
\begin{equation}
\label{eq:E1}
	\hat H_{E1} = - {\bf d \cdot E}(e^{-i\omega t}+e^{i\omega t}),
\end{equation}
where ${\bf d}$ is the dipole moment operator.

In the TDHF method, a single-electron wave function is
\begin{equation}
\label{eq:wf}
	\psi = \psi_0 + \delta \psi + Xe^{-i\omega t}+Ye^{i\omega t}+
	\delta Xe^{-i\omega t}+ \delta Ye^{i\omega t},
\end{equation}
where $\delta \psi$ is the correction due to the weak interaction
acting alone, $X$ and $Y$ are corrections due to the photon field
acting alone, and $\delta X$ and $\delta Y$ are corrections due
to both fields acting simultaneously.
These corrections are found by solving self-consistently the system
of the TDHF equations for the core states
\begin{eqnarray}
	(\hat{H}_{0}-\epsilon)\delta \psi &=&
	-(\hat{H}_{W}+\delta \hat{V}_{W})\psi, \label{eq:WE1:1}\\
	(\hat{H}_{0}-\epsilon -\omega)X &=&
	-(\hat{H}_{E1}+\delta \hat{V}_{E1})\psi, \label{eq:WE1:2}\\
	(\hat{H}_{0}-\epsilon +\omega)Y&=&
	-(\hat{H}_{E1}^{\dagger}+\delta \hat{V}_{E1}^{\dagger})\psi,
	\label{eq:WE1:3}\\
	(\hat{H}_{0}-\epsilon -\omega)\delta X &=&
	-\delta \hat{V}_{E1}\delta \psi
	-\delta \hat{V}_{W}X
	-\delta \hat{V}_{E1W}\psi, \label{eq:WE1:4}\\
	(\hat{H}_{0}-\epsilon +\omega)\delta Y&=&
	-\delta \hat{V}_{E1}^{\dagger}\delta \psi
	-\delta \hat{V}_{W}^{\dagger}Y
	-\delta \hat{V}_{E1W}^{\dagger} \psi,\label{eq:WE1:5}
\end{eqnarray}
where $\delta \hat{V}_W$ and $\delta \hat{V}_{E1}$ are corrections to the
core potential due to the weak and E1 interactions, respectively,
and $\delta \hat{V}_{E1W}$ is the correction to the core potential
due to the simultaneous action of the weak field and the electric field
of the photon.

The TDHF contribution to $E_{PNC}$ between the states $6s$ and $7s$
is given by
\begin{equation}
\label{eq:TDHF}
E_{PNC}^{TDHF}=\langle \psi _{7s}|\hat{H}_{E1}+\delta \hat{V}_{E1}|
\delta \psi _{6s}\rangle +
\langle \psi _{7s}|\hat{H}_{W}+\delta \hat{V}_{W}|
X _{6s}\rangle +
\langle \psi _{7s}|\delta \hat{V}_{E1W}|\psi _{6s}\rangle  \ .
\end{equation}
The corrections $\delta \psi_{6s}$ and $X_{6s}$ are found by solving
the equations~(\ref{eq:WE1:1}-\ref{eq:WE1:2}) in the field of the
frozen core (of course, the amplitude (\ref{eq:TDHF}) can instead
be expressed in terms of corrections to $\psi _{7s}$).

\begin{figure}[b]
\centerline{\psfig{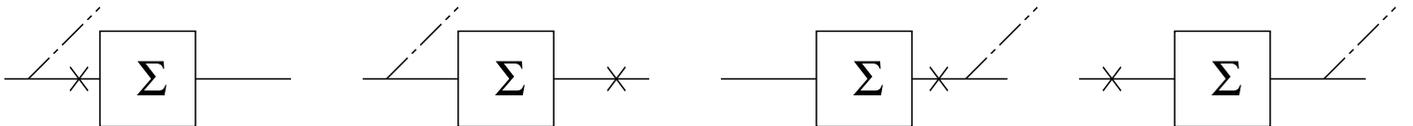}}
\caption{Lowest-order correlation corrections to the PNC E1 transition
amplitude. Dashed line is the E1 field; cross is the nuclear weak field.}
\label{fig:pncdom}
\end{figure}

If we use Brueckner orbitals instead of RHF orbitals to calculate
the PNC amplitude in Eq.~(\ref{eq:TDHF}) we include all-orders
in $\hat{\Sigma}$ contributions to $E_{PNC}$.
However, the correlation potential is energy-dependent,
$\hat{\Sigma}=\hat{\Sigma}(\epsilon)$, which means that $\hat \Sigma$
operators for the $6s$ and $7s$ states are different.
We should consider the proper energy-dependence at least in first-order
in $\hat \Sigma$ (higher-order corrections are small and the proper
energy-dependence is not important for them).
The first-order in $\hat{\Sigma}$ corrections to $E_{PNC}$ are
presented diagrammatically in Fig.~\ref{fig:pncdom}.
We can write these as
\begin{equation}
\label{eq:pnc-cor}
\langle \psi _{7s}|\hat{\Sigma}_s(\epsilon_{7s})|\delta X_{6s}\rangle
+\langle \delta\psi _{7s}|\hat{\Sigma}_p(\epsilon_{7s})|X_{6s}\rangle
+\langle \delta Y_{7s}|\hat{\Sigma}_s(\epsilon_{6s})|\psi_{6s}\rangle
+\langle Y_{7s}|\hat{\Sigma}_p(\epsilon_{6s})|\delta \psi_{6s}\rangle \ .
\end{equation}
The non-linear in $\hat{\Sigma}$ contribution can be found by subtracting
from the all-orders result the first-order value found in the same method.

\begin{figure}[b]
\centerline{\psfig{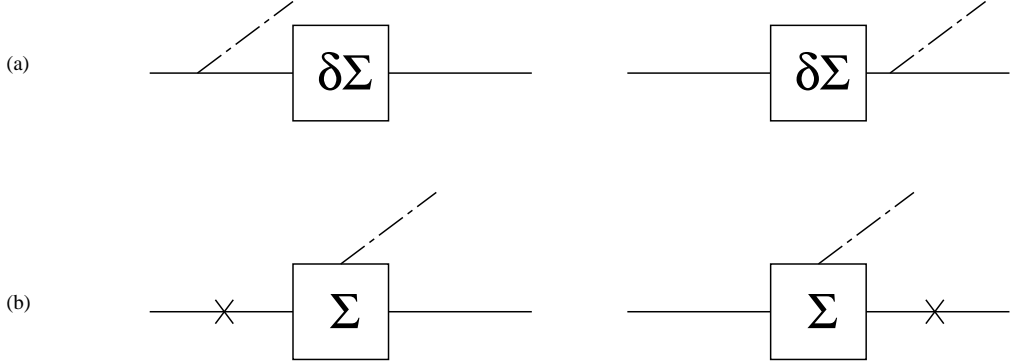}}
\caption{Small corrections to the PNC E1 transition amplitude:
external field inside the correlation potential.
In diagrams (a) the weak interaction is inside the
correlation potential ($\delta \hat{\Sigma}$ denotes the change in
$\hat{\Sigma}$ due to the weak interaction);
this is known as the weak correlation potential.
Diagrams (b) represent structural radiation
(photon field inside the correlation potential).}
\label{fig:pncint}
\end{figure}

The correlation corrections to $E_{PNC}$ we have considered so far
are usually called ``Brueckner-type'' corrections.
(In this case the external field interacts with the external electron
lines.) There are also contributions to $E_{PNC}$ in which the
external field acts inside the correlation potential
(see Fig.~\ref{fig:pncint}).
Those diagrams in which the E1 interaction occurs in the internal
lines are known as ``structural radiation'',
while those in which the weak interaction occurs in the internal lines are
known as the ``weak correlation potential''.
There is another second-order correction to the amplitudes which arises
from the normalization of states \cite{dzuba87}.
The structural radiation, weak correlation potential, and
normalization contributions are suppressed by the small parameter
$E_{\rm ext}/E_{\rm core}\sim 1/10$, where  $E_{\rm ext}$ and
$E_{\rm int}$ are excitation energies of the external and core electrons,
respectively.

We have also included a correction to $E_{PNC}$ due to the Breit interaction.
We calculated this in a way similar to the earlier work \cite{harabati}.
The main difference is that in the present calculations we have also
included the Breit contribution to the last term in Eq.~(\ref{eq:TDHF}).
This makes the calculations more consistent but doesn't change the result
significantly.

We postpone the analysis of radiative corrections until
Section~\ref{sec:rad}.

\section{$6s-7s$ PNC amplitude}
\label{sec:results}

The results of our calculation for the  $6s-7s$ PNC amplitude
are presented in Table~\ref{tab:pnci}.
Notice that the time-dependent Hartree-Fock value gives a
contribution to the total amplitude of about $98\%$.
The point is that there is a strong cancellation of the
correlation corrections to the PNC amplitude.
The stability of the PNC amplitude compared to other
quantities in which the correlation corrections are large
will be discussed in more detail in Section~\ref{sec:accuracy}.
Notice that the values do not differ significantly from our 1989 results
(compare the 1989 final result $0.908\times 10^{-11}iea_{B}(-Q_{W}/N)$
with ``Subtotal'' of Table~\ref{tab:pnci} for the current calculation).
The new series of higher-order diagrams and higher numerical accuracy
of the current work has therefore not changed the previous result
(note, however, that particular contributions are slightly different).

The mixed-states approach has also been performed in
\cite{blundell90} and \cite{kozlov01} to determine the PNC amplitude
in cesium.
However, in these works the screening of the
electron-electron interaction was included in a simplified way.
In \cite{blundell90} empirical screening factors were placed before
the second-order correlation corrections $\hat{\Sigma}^{(2)}$ to fit the
experimental values of energies.
Kozlov {\it et al.} \cite{kozlov01} introduced screening factors
based on average screening factors calculated for the Coulomb
integrals between valence electron states.
The results obtained by these groups
(without the Breit interaction, i.e., corresponding to the
Subtotal of Table~\ref{tab:pnci})
are $0.904$ \cite{blundell90} and $0.905$ \cite{kozlov01}.
To be sure that we understand the difference between these values and
our value, we performed a pure second-order
(i.e., using $\hat{\Sigma}^{(2)}$) calculation and fitted the energies
(as was done in \cite{blundell90}) and reproduced their result, $0.904$.

In the work \cite{blundell90} a calculation using the
sum-over-states method was also performed.
In the sum-over-states approach the $6s-7s$ PNC amplitude
is expressed in the form
\begin{equation} \label{sum}
E_{PNC}=\sum _{n} \Big(
\frac{\langle 7s |\hat{H}_{W}|np\rangle \langle np|\hat{H}_{E1}|6s \rangle}
{E_{7s}-E_{np}} +
\frac{\langle 7s |\hat{H}_{E1}|np\rangle \langle np|\hat{H}_{W}|6s \rangle}
{E_{6s}-E_{np}} \Big) \ .
\end{equation}
The authors of reference \cite{blundell90} include single, double, and
selected triple excitations into their wave functions.
Note, however, that even if wave functions of $6s$, $7s$, and intermediate
$np$ states are calculated exactly
(i.e., with all configuration mixing included)
there are still some missed contributions in this approach.
Consider, e.g., the intermediate state $6p\equiv 5p^{6}6p$.
It contains an admixture of states $5p^{5}ns6d$:
$\widetilde{6p}=5p^{6}6p + \alpha 5p^{5}ns6d+...$.
This mixed state is included into the  sum (\ref{sum}).
However, the sum (\ref{sum}) must include all many-body states
of opposite parity.
This means that the state $\widetilde{5p^{5}ns6d}=
5p^{5}ns6d- \alpha 5p^{6}6p+...$ should also be included into
the sum. Such contributions to $E_{PNC}$ have never
been estimated directly within the sum-over-states approach.
However, they are included into our mixed-states calculation.
The result of the sum-over-states approach, 0.909,
is very close to the result of the mixed-states approach, 0.908.
It is important to note that the omitted higher-order many-body corrections
are different in these two methods.
This may be considered as an argument that the omitted many-body corrections
in both calculations are small.
Of course, here we assume that the omitted many-body corrections to both
values (which, in principle, are completely different) do not
``conspire'' to give exactly the same magnitude.

Therefore we will take $0.908$ for the value of $E_{PNC}$
(Subtotal of Table~\ref{tab:pnci}) as this corresponds
to the most complete mixed-states calculation
and is in agreement with the sum-over-states calculation of
reference \cite{blundell90}.

With Breit, our result becomes $0.902\times 10^{-11}iea_{B}(-Q_{W}/N)$.
This correction is in agreement with
\cite{derevianko00,harabati,kozlov01}.

We use the two-parameter Fermi model for the proton and neutron distributions:
\begin{equation}
\rho (r)=\rho _{0}\Big[ 1+\exp [(r-c)/a] \Big] ^{-1}\ ,
\end{equation}
where $t=a(4\ln 3)$ is the skin-thickness,
$c$ is the half-density radius, and
$\rho _{0}$ is found from the normalization condition $\int \rho dV=1$.
In 1989 the thickness and half-density radius for the proton
distribution were taken to be $t_{p}=2.5~{\rm fm}$ and
$c_{p}=5.6149~{\rm fm}$ (corresponding to a root-mean-square
(rms) radius $\langle r_{p}^{2}\rangle ^{1/2}=4.836~{\rm fm}$).
In this work we have used improved parameters
$t_{p}=2.3~{\rm fm}$ and $c_{p}=5.6710~{\rm fm}$
($\langle r_{p}^{2}\rangle ^{1/2}=4.804~{\rm fm}$) \cite{fricke95}.
This changes the wave functions slightly,
leading to a very small correction to the PNC amplitude
of $0.08\%$ ($0.0007$).
(This is in agreement with a simple analytical estimate:
the factor accounting for the change in the electron density
is $\sim (4.804/4.836)^{-Z^{2}\alpha ^{2}}\sim 0.1\% \ $.)
This correction has already been included into the TDHF value.

In the work \cite{dzuba89} we used the proton distribution in the
weak interaction Hamiltonian (Eq.~(\ref{eq:h_w})).
In the current work we have found the small correction to
$E_{PNC}$ which arises from taking the (poorly understood)
neutron density in Eq.~(\ref{eq:h_w}).
We use the result of Ref. \cite{r_{np}} for the difference
$\Delta r_{np}=0.13(4)~{\rm fm}$
in the root-mean-square radii of the neutrons
$\langle r_{n}^{2}\rangle ^{1/2}$ and protons
$\langle r_{p}^{2}\rangle ^{1/2}$.
We have considered three cases which correspond to the same value of
$\langle r_{n}^{2}\rangle$: (i) $c_{n}=c_{p}$, $a_{n}>a_{p}$;
(ii) $c_{n}>c_{p}$, $a_{n}>a_{p}$; and (iii) $c_{n}>c_{p}$, $a_{n}=a_{p}$
(using the relation $\langle r_{n}^{2}\rangle \approx \frac{3}{5}c_{n}^{2}
+\frac{7}{5}\pi ^{2}a_{n}^{2}$).
We have found that $E_{PNC}$ shifts from $-0.18\%$ to $-0.21\%$
when moving from the extreme $c_{n}=c_{p}$ to the extreme $a_{n}=a_{p}$.
Therefore, $E_{PNC}$ changes by about $-0.2\%$ ($-0.0018$)
due to consideration of the neutron distribution.
This is in agreement with Derevianko's estimate,
$-0.19(8)\%$ \cite{derevianko}.

In the next section we discuss the radiative corrections to
$E_{PNC}$. The dominating contribution comes from the Uehling potential and
increases the amplitude by 0.4\%.

Therefore, we have
\begin{equation}
E_{PNC}=0.9041 \times 10^{-11}iea_{B}(-Q_{W}/N)
\end{equation}
as our central point for the PNC amplitude.
The error will be estimated in Section~\ref{sec:accuracy}.

\section{QED-type radiative corrections to energy levels,
wave functions, and the PNC amplitude}
\label{sec:rad}

The radiative corrections to the weak charge $Q_W$ have been
calculated for the free electron. However, an electron in a heavy
atom is bound, and this produces additional radiative
corrections proportional to $\alpha (Z \alpha)^n$,
$n=1,2,...$. Recently such corrections were considered
by Milstein and Sushkov \cite{milstein}. They found that
the most important are corrections enhanced by
the large parameter $\ln(\lambda/R)$, where $\lambda=\hbar/mc$
is the electron Compton wavelength and $R$ is the nuclear
radius. This type of correction arises from the radiative
corrections to the electron wave function near the nucleus.
In this region the $s$-wave and $p_{1/2}$-wave (lower Dirac component)
electron densities are singular, $|\psi(r)|^2 \sim r^{-Z^2\alpha^2}$.
The radiative corrections modify the potential at small distances
$r<\lambda$, ${\tilde V }(r)= -Z \alpha (1+\delta)/r$.
Correspondingly, the electron wave functions change,
$|\psi(r)|^2 \sim r^{-Z^2\alpha^2 (1 +\delta)^2}$ for $r<\lambda$.
This gives the radiative correction factor for the electron
density inside the nucleus,
\begin{equation}
\label{psirad}
\frac{|\psi(R)|^2}{|\psi(\lambda)|^2} \sim \Big( \frac{\lambda}{R}\Big)
^{Z^2\alpha^2 2\delta}=\exp \Big( 2\delta Z^2\alpha^2 \ln(\lambda /R) \Big)
\ .
\end{equation}
For the Uehling (vacuum polarization) potential
$\delta \sim \alpha \ln(\lambda /r)$ \cite{berestetskii}.
This gives an additional power of the large  parameter $\ln(\lambda / R)$.
This led Milstein and Sushkov \cite{milstein} to conclude that the
Uehling potential gives the dominating radiative correction
to $E_{PNC}$, $\sim  Z^2\alpha^3 \ln^2(\lambda/ R)$.
Numerical calculations of the Uehling potential contribution
have been performed in \cite{johnson01} and in the present work.
This radiative correction increases $E_{PNC}$ by 0.4\%.

Milstein and Sushkov \cite{milstein} demonstrated that there are no other
radiative corrections which are enhanced by  $\ln^2(\lambda /R)$.
However, any correction to the potential with nonzero
$\delta(R) \sim \alpha$ gives a correction to the electron density
$\sim  Z^2\alpha^3 \ln (\lambda /R)$. There are also corrections
$\sim  Z^2\alpha^3 \ln (Z^2\alpha^2)$ which originate from the
shift of the energy levels (Lamb shift). We briefly discuss
these corrections below.

Let us start our discussion from the radiative corrections to energy
levels.
The calculation of the shift can be divided into two parts:
one in which the electron interaction with virtual photons of
high-frequency are considered, and one in which
virtual photons of low-frequency are considered.

\begin{figure}[b]
\centerline{\psfig{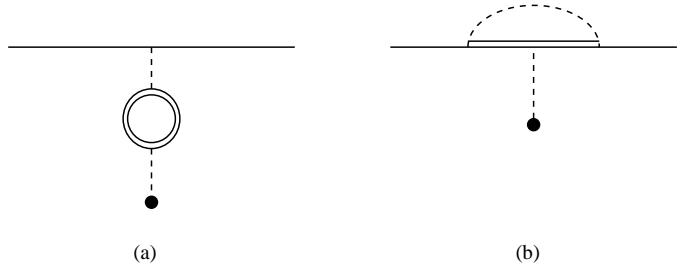}}
\caption{High-frequency contribution to radiative corrections.
Diagram (a) corresponds to the Uehling potential.
Diagram (b) is the vertex correction.
The single solid line represents the bound electron;
the double line is the free electron;
the Coulomb interaction is denoted by the dashed line;
and the filled circle denotes the nucleus.}
\label{fig:rad}
\end{figure}

In the high-frequency case the external field
(the strong nuclear Coulomb field) need only be included
to first-order.
In this case the contributions to the Lamb shift arise
from the diagrams presented in Fig.~\ref{fig:rad}.
The contribution of the Uehling potential (Fig.~\ref{fig:rad}(a))
to the Lamb shift is very small.
The main contribution comes from the vertex correction
(Fig.~\ref{fig:rad}(b)).
(In the case of a free electron the vertex diagrams give the electric
$f(q^{2})$ and magnetic $g(q^{2})$ form factors.)
The perturbation theory expression for $f(q^2)$ contains an
infra-red divergence and requires a low-frequency
cut-off parameter $\kappa$ - see, e.g., \cite{berestetskii}.
Assuming $q^2 \ll m^2 c^2$, the high-frequency contribution to the
Lamb shift can be presented as a potential given by the following
expression \cite{berestetskii}
\begin{eqnarray}
\delta \Phi ({\bf r})
&=&\Big[ \delta \Phi _{f}+\delta \Phi _{U}\Big] +\delta \Phi _{g}\nonumber \\
&=&\frac{\alpha \hbar ^{2}}{3\pi m^{2}c^{2}}\Big(
\ln \frac{m}{2\kappa} +\frac{11}{24}-\frac{1}{5}\Big)
\Delta \Phi ({\bf r})-
i\frac{\alpha \hbar}{4\pi mc}\mbox{\boldmath$\gamma$}\cdot
\mbox{\boldmath$\nabla$}\Phi ({\bf r}) \ .
\end{eqnarray}
For the Coulomb potential, $\Delta \Phi =-4\pi Ze\delta ({\bf r})$.
Here the last long-range term ($\delta \Phi _{g}$) comes from
the anomalous electron magnetic moment ($g(0)$);
the infra-red cut-off parameter $\kappa$ appears from the electric
form factor $f(q^{2})$.
This term with the large $\ln \frac{m}{2\kappa}$ gives the dominant
contribution to the Lamb shift of $s$-levels.
The infra-red divergence for $\kappa \rightarrow 0$ indicates the
importance of the low-frequency contribution for this term.

If we go beyond the approximation
$q^{2}<<m^{2}c^{2}$, the term $\delta \Phi _{f}$ should be
associated with a non-local self-energy operator
${\hat \Sigma}_{\rm rad} ({\bf r},{\bf r}',E)$ with typical values
$|{\bf r}-{\bf r}'|\lesssim \frac{\hbar}{mc}$
and $r\sim r' \lesssim \frac{\hbar}{mc}$.
We need this operator in a simple limit, $E<<mc^{2}$.
Matrix elements of ${\hat \Sigma}_{\rm rad} ({\bf r},{\bf r}',0)$
depend only on the electron density near the origin. Therefore,
we can express the Lamb shift of the external electron state in a
neutral atom in terms of the known Lamb shift
of highly excited states in hydrogen-like ions. To implement this
scheme  we can approximate
 ${\hat \Sigma}_{\rm rad} ({\bf r},{\bf r}',0)$
 by a two-parametric $(A,b)$ potential
\begin{equation} \label{param}
\delta \Phi _{f}=-A\frac{\alpha}{\pi} r {\rm e}^{-b\frac{mc}{\hbar}r}\Phi \ ,
\end{equation}
where the factor $r$ is introduced to remove the singularity at $r=0$.
(We have also performed the calculation using the potential
$\delta \Phi _{f}$ without this factor $r$.
The results for the energy levels are the same.)
The parameters $A$ and $b$ in $\delta \Phi _{f}$ can be found from the
fit of the Lamb-shift of the high Coulomb levels $3s$, $4s$, $5s$ and
$3p$, $4p$ and $5p$ (in one-electron ions) which were calculated as
a function of the nuclear charge $Z$ in Refs. \cite{mohr}.
We have checked that $A_s=180$, $A_p=90$ and $b=1$ fit all these
Lamb shifts quite accurately
(for the potential without the factor $r$ we have found
$A_s=1.17$ and $A_p=1.33$).
The anomalous magnetic moment contribution is
\begin{equation}
\delta \Phi _{g}=-i\frac{\alpha \hbar}{4\pi mc}\mbox{\boldmath$\gamma$}
\cdot \mbox{\boldmath$\nabla$}\Phi \ .
\end{equation}
The Uehling potential for a finite nucleus is given by \cite{fullerton76}
(in atomic units: $\hbar =m=e=1$, $\alpha =1/c$)
\begin{equation}
\delta \Phi _{U}=-\frac{2\alpha ^{2}}{3r}\int _{0}^{\infty}dx~
x\rho(x)\int _{1}^{\infty}dt~\sqrt{t^{2}-1}
\Big(
\frac{1}{t^{3}}+\frac{1}{2t^{5}}\Big)
\Big(
{\rm e}^{-2t|r-x|/\alpha} -{\rm e}^{-2t(r+x)/\alpha} \Big)    \ ,
\end{equation}
where $\rho (x)$ is the nuclear charge density.
It is more convenient to use a simpler formula for
$\delta \Phi _{U}$ for $r\geq R$,
\begin{eqnarray}
&&\delta \Phi _{U}(r)= \Phi (r)\frac{\alpha ^{4}}{8\pi R^{3}}
\int _{1}^{\infty}dt~ \sqrt{t^{2}-1}
\Big(
\frac{1}{t^{5}}+\frac{1}{2t^{7}}\Big)
{\rm e}^\frac{-2tr}{\alpha}I(x)  \ , \\
&&I(x)=-{\rm e}^{x}+{\rm e}^{-x}+x{\rm e}^{x} +x{\rm e}^{-x} \ ,
\qquad x=2tR/\alpha \ ,
\end{eqnarray}
and take
$\frac{\delta \Phi _{U}(r<R)}{\Phi (r<R)}=
\frac{\delta \Phi _{U}(r=R)}{\Phi (r=R)}$.
There is practically no loss of numerical accuracy in this
approximation since a typical scale for the variation of
$\delta \Phi _{U}/\Phi$ is given by the electron Compton length
$\frac{\hbar}{mc}>>R$.

Since  $Z\alpha=0.4 $ is not so small for Cs it is important to estimate
the contribution of the higher-order in $Z\alpha$ vacuum-polarization
correction (the Wichmann-Kroll term \cite{kroll}). For simplicity we use
an approximate expression for this potential
\begin{equation}
\label{eq:WK}
	\delta \Phi _{WK} = -\frac{2}{3}\frac{\alpha}{\pi} \,
	\frac{0.092 Z^2 \alpha^2}{1+\big(\frac{1.62r}{\alpha}\big)^4}\Phi
\end{equation}
which is exact at small and large distances (a small-$r$ asymptotic was
presented in Refs. \cite{milstein83,milstein}).
We have found that the contribution of this potential to the $s$-wave
energies is $\sim 30$ times smaller than that of the Uehling potential.
The contribution of the Uehling potential $\delta \Phi _{U}$ to the Lamb
shift is always small (less than 15\%). Note that the contribution
of the radiative corrections to the electron core electrostatic
potential can be estimated using the semiclassical expression
for  ${\hat \Sigma}_{\rm rad} ({\bf r},{\bf r}',E)$
presented in Ref. \cite{zelevinsky}. This contribution
for $s$-levels is two orders of magnitude smaller than
that of the nuclear potential
(for higher angular momenta the nuclear contribution is small and
the electron contribution is relatively important).

The potential $\delta \Phi _{g}$ due to the magnetic form factor
is a long-range one.
This also hints that there are no large higher $Z\alpha$
corrections here. The contribution of this potential to the Lamb shift
is about 30\% for $s$-levels and 70\% for $p$-levels.
Note that for the term $\delta \Phi _{f}$ and the total Lamb-shift
we do not use the assumption $Z\alpha <<1$ since we fitted the exact results
for the single-electron ions.

The radiative corrections for Cs energy levels are presented
in Table~\ref{tab:radenergies}.

Now we  discuss the contribution of
QED-type radiative corrections to the electron wave function
and  PNC amplitude $E_{PNC}$.
Note that it is not enough to calculate the radiative corrections to
the matrix element of the weak interaction
$\langle n'p_{1/2}|\hat{H}_{W}|ns\rangle$.
Corrections to the energy intervals like $6s-6p$  are also important
since these intervals are small at the scale of the atomic unit
($ \sim 1/20$) and sensitive to perturbations.
The change in the energies also influences the large-distance behavior
of the electron wave functions which determine the usual
E1 amplitudes in the sum-over-states approach.
If we keep only three dominating terms in the sum-over-states approach
(see below) the contributions of the energy shifts (-0.29 \%)
and corrections to the amplitudes (0.33\%) cancel each other.
The simplest way to find the total answer (including
the corrections to the weak matrix elements) is to include $\delta \Phi$
into the relativistic Hartree-Fock equations and
then perform all calculations.

The results are the following:
the largest contribution to $E_{PNC}$ comes from the Uehling
potential $\delta \Phi _{U}$. It increases $E_{PNC}$ by
$0.41\%$ (in agreement with \cite{johnson01,milstein}).
The contribution of $\delta \Phi _{g}$ is  $-0.03 \%$
(due to cancellation of the contributions of the corrections
to the $s$-wave and $p$-wave); the Wichmann-Kroll contribution is
$-0.006\%$. Note that the latter contribution can be estimated
analytically using Eq.~(\ref{psirad}). The ratio of the Wichmann-Kroll
contribution to the Uehling contribution in the logarithmic approximation is
$-0.184~Z^2\alpha^2/\ln (\lambda /R)$.

The contribution of $\delta \Phi _{f}$ is well-defined only when
it originates from large distances (due to the shift of the energy levels
and the large-distance behavior of the electronic wave function
which influences electromagnetic amplitudes). A small-distance
contribution is not gauge-invariant and should be treated simultaneously
with the correction to the Z-boson exchange vertex
\cite{milstein}. Therefore, we deliberately selected a non-singular
parametric potential (\ref{param}) to approximate $\delta \Phi _{f}$.
This potential does not generate any singular
(in nuclear radius $R$)  contributions proportional to the large
parameter $ \ln (\lambda /R)$. This means that the subject
of our discussion now is different from the radiative corrections
 to the weak matrix element considered by Milstein and Sushkov
in \cite{milstein}. They were interested in contributions
enhanced by this large parameter $ \ln (\lambda /R)$.

The main contribution to the Lamb shift comes from the distances
$r < 1/Z$ (in atomic units). This corresponds to the parameter
$b>Z\alpha $ in the potential (\ref{param}). On the other hand
a typical distance for the radiative corrections
is $r \sim \lambda=\frac{\hbar}{mc}$. Therefore, to estimate
the contribution of $\delta \Phi _{f}$ we performed calculations
for two extreme values of the parameters $b=1$ and $b=Z\alpha $
(the values of the Lamb shifts are kept the same in both cases
by selecting appropriate values of the parameter $A$).
For $b=1 $ this contribution is $-0.2\%$. For  $b=Z\alpha $
it is -0.08\%. This gives us an indication that the large-distance
contribution to the radiative corrections to
$E_{PNC}$ is small. On the other hand, according to Sushkov and
Milstein \cite{milstein}, the small-distance
contribution to $E_{PNC}$ is dominated by the Uehling
potential which was considered above.

\section{Estimate of accuracy of PNC amplitude}
\label{sec:accuracy}

We have estimated the error of the PNC amplitude in a number of different
ways. There are two main methods:
(i) root-mean-square (rms) deviation of the calculated energy intervals,
E1 amplitudes, and hyperfine structure constants
from the accurate experimental values;
(ii) influence of fitting of energies and hyperfine structure
constants on the PNC amplitude.

\subsection{Root-mean-square deviation}

Remember that the PNC amplitude can be expressed as a sum over
intermediate states (see formula~(\ref{sum})).
Each term in the sum is a product of E1 transition amplitudes,
weak matrix elements, and energy denominators.
There are three dominating contributions to the $6s-7s$ PNC amplitude
in Cs:
\begin{eqnarray}
E_{PNC}&=&
\label{sumcs}
\frac{\langle 7s|\hat{H}_{E1}|6p\rangle \langle 6p|\hat{H}_{W}|6s\rangle}
{E_{6s}-E_{6p}} +
\frac{\langle 7s|\hat{H}_{W}|6p\rangle \langle 6p|\hat{H}_{E1}|6s\rangle}
{E_{7s}-E_{6p}}+
\frac{\langle 7s|\hat{H}_{E1}|7p\rangle \langle 7p|\hat{H}_{W}|6s\rangle}
{E_{6s}-E_{7p}}+...
\nonumber \\
&=& -1.908+1.493+1.352+...=0.937 +... \
\end{eqnarray}
(the numbers are from the work \cite{blundell90}
where the sum-over-states method was used;
here we just demonstrate that these terms dominate).
While we do not use the sum-over-states approach in our calculation of
the PNC amplitude, it is instructive to analyze the accuracy of
the E1 transition amplitudes, weak matrix elements, and
energy intervals which contribute to Eq.~(\ref{sumcs}) as
they have been calculated using the same method as that used to
calculate $E_{PNC}$.

Let us begin with the energy intervals.
The calculated removal energies are presented in Table~\ref{tab:energies}.
The Hartree-Fock values deviate from experiment by $10\%$.
Including the second-order correlation corrections ${\hat \Sigma} ^{(2)}$
reduces the error to $\sim 1\%$.
When screening and the hole-particle interaction are included into
${\hat \Sigma} ^{(2)}$ in all orders,
the energies improve, $\sim 0.3\%$.
The percentage deviations from experiment of the energy intervals of interest
are: $E_{6s}-E_{6p}$, $0.3$; $E_{7s}-E_{6p}$, $0.4$;
and $E_{6s}-E_{7p}$, $0.3$. The rms error is $0.3\%$.
We can in fact reproduce energy intervals exactly by placing coefficients
before the correlation potential. We will use this procedure as another
test of the stability of the results. Note, however, that the accuracy for
the energies is already very high and the remaining discrepancy with
experiment is of the same order of magnitude as the Breit and radiative
corrections. Therefore, generally speaking, we should not expect that
fitting of the energy will always improve the results for amplitudes and
hyperfine structure. In fact, as we will see below, some values do improve
while others do not. The overall accuracy, however, remains at the same level.

The relevant E1 transition amplitudes (radial integrals) are presented in
Table~\ref{tab:e1i}.
These were calculated with the energy-fitted ``bare'' correlation
potential $\hat{\Sigma}^{(2)}$ and the (unfitted and fitted)
``dressed'' potential $\hat{\Sigma}$.
Structural radiation and normalization contributions were also
included.
In Table~\ref{tab:e1ii} the percentage deviations of
the calculated values from experiment are listed.
Without energy fitting, the rms error is $0.1\%$.
Fitting the energy gives a rms error of $0.2\%$ for $\hat{\Sigma}^{(2)}$
and $0.3\%$ for the complete $\hat{\Sigma}$.

We cannot directly compare weak matrix elements with experiment.
Like the weak matrix elements, hyperfine structure is determined
by the electron wave functions in the vicinity of the nucleus,
and this is known very accurately.
The hyperfine structure constants calculated in different approximations
are presented in Table~\ref{tab:hfsi}.
Corrections due to the Breit interaction, structural radiation,  and
normalization are included.
The percentage deviations from experiment are shown in
Table~\ref{tab:hfsii}.
The rms deviation of the calculated hfs values from experiment
using the unfitted ${\hat \Sigma}$ is $0.5\%$.
With fitting, the rms error in the pure second-order approximation is $0.3\%$;
with higher orders we get $0.4\%$.
We are, however, trying to estimate the accuracy of the $s-p$ weak
matrix elements. It seems reasonable for us to use the square-root
formula, $\sqrt{{\rm hfs}(s){\rm hfs}(p)}$.
The errors are presented in Table~\ref{tab:hfsii}.
Notice that by using this approach the error is smaller.
Without energy fitting, the rms error is $0.5\%$.
With fitting, the rms error in the second-order calculation
($\hat{\Sigma} ^{(2)}$) is $0.2\%$ and in the full calculation
($\hat{\Sigma}$) it is $0.3\%$.

From this section we can conclude that the rms error for the relevant
parameters is $0.5\%$ or better.

Note that from this analysis the error for the sum-over-states
calculation of $E_{PNC}$ would be larger than this, as
the errors for the energies, hfs constants, and E1 amplitudes
contribute to each of the three terms in Eq.~(\ref{sumcs}).
However, in the mixed-states approach, the errors do not add in this
way.
We get a better indication of the error of our calculation
of $E_{PNC}$ in the next section.

\subsection{Influence of fitting on the PNC amplitude}

In the section above we presented calculations in three different
approximations:
with unfitted $\hat{\Sigma}$,
and with $\hat{\Sigma}^{(2)}$ and $\hat{\Sigma}$ fitted with
coefficients to reproduce experimental removal energies.
The errors in these approximations are of different magnitudes and signs.
We now calculate the PNC amplitude using these three approximations.
The spread of the results can be used to estimate the error.

The results are listed in Table~\ref{tab:pncii}.
It can be seen that the PNC amplitude is very stable.
The PNC amplitude is much more stable than hyperfine structure.
This can be explained by the much smaller correlation corrections
to $E_{PNC}$ ($\sim 2\%$ for $E_{PNC}$ and $\sim 30\%$ for hfs;
compare Table~\ref{tab:pnci} with Table~\ref{tab:hfsi}).
One can say that this small value of the correlation correction is
a result of cancellation of different terms in (\ref{eq:pnc-cor})
but each term is not small (see Table~\ref{tab:pnci}). However,
this cancellation has a regular behavior. The same correlation
potential $\hat \Sigma$ is used to calculate energies and correlation
corrections (\ref{eq:pnc-cor}) to $E_{PNC}$. Therefore, whatever
way is used to calculate $\hat \Sigma$, if the accuracy for energies
is good, the correlation correction (\ref{eq:pnc-cor}) is stable.
The stability of $E_{PNC}$ may be compared to the stability of the
usual electromagnetic amplitudes where the error is very small
(even without fitting).

We have also considered the fitting of hyperfine structure using
different coefficients before each $\hat{\Sigma}$.
The first-order in $\hat{\Sigma}$ correlation correction (\ref{eq:pnc-cor})
changes by about $10\%$. This changes the PNC amplitude
by about $0.4\%$.

\vspace{5mm}

It is also instructive to look at the spread of $E_{PNC}$ obtained
in different schemes.
(This has already been discussed in some detail in
Section~\ref{sec:results}.)
The result of the present work is in excellent agreement with our earlier
result \cite{dzuba89} while the calculation scheme is significantly different.
The only other calculation of the $E_{PNC}$ in Cs which is as
complete as ours is that of Blundell {\it et al.} \cite{blundell90}.
Their result in the all-orders sum-over-states approach is 0.909
(without Breit) and is very close to our value of 0.908
(corresponding to ``Subtotal'' of Table~\ref{tab:pnci}).
Our result (0.904) obtained in second-order with fitting of the energies
is useful in determining the accuracy of the calculations of $E_{PNC}$.
(Remember that this value is in agreement with results of similar
calculations performed in \cite{blundell90,kozlov01}; see
Section~\ref{sec:results}.) One can see that replacing
the all-order $\hat \Sigma$ by its very rough second-order (with fitting)
approximation changes $E_{PNC}$ by less than 0.4\% only. On the other hand,
if the higher orders are included accurately, the difference between the
two very different approaches is 0.1\% only.

The maximum deviation we have obtained in the above analysis is
$0.5\%$. We will therefore use this as our estimate for the
uncertainty of the $E_{PNC}$ calculation.

We do not include the error associated with the radiative corrections
into the estimate of the accuracy since the large-distance
($r>\lambda$) contribution of the radiative corrections is small
and the small-distance contribution ($R\lesssim r<\lambda$)
enhanced by $\ln (\lambda/R)$ can be accurately calculated.

\section{Conclusion}

We have obtained the result
\begin{equation}
E_{PNC}=0.904\Big(1\pm 0.5\%\Big) \times 10^{-11}iea_{B}(-Q_{W}/N)
\end{equation}
for our calculation of the $6s-7s$ PNC amplitude in Cs.
This is in agreement with other PNC calculations,
however we would like to emphasize that our calculation is
the most complete.
The most precise measurement of the $6s-7s$ PNC amplitude in Cs
is \cite{wood97}
\begin{equation}
-\frac{{\rm Im} (E_{PNC})}{\beta}=1.5939(56)\frac{\rm mV}{\rm cm} \ ,
\end{equation}
where $\beta$ is the vector transition polarizability.
There are currently two very precise values for $\beta$. One value
\begin{equation}
\label{eq:beta1}
\beta =26.957(51) a_{B}^{3}
\end{equation}
was obtained in our analysis \cite{dzuba00} of the Bennett and Wieman
measurements \cite{wieman99} of the
$M1_{\rm hfs}/\beta$ ratio \cite{bouchiat88}.
We have obtained another value
\begin{equation}
\label{eq:beta2}
\beta =27.15(11) a_{B}^{3}
\end{equation}
from the measurement \cite{cho97} of $\alpha/\beta$ and
an analysis, similar to that of Ref. \cite{dzuba97}, using
the most accurate experimental data for the E1 transition amplitudes
including recent measurements of Vasilyev {\it et al.} \cite{vasilyev01}.
The errors in Eqs.~(\ref{eq:beta1},\ref{eq:beta2}) are obtained by adding
in quadrature the experimental and theoretical errors.
Notice that the central point of our value (Eq.~(\ref{eq:beta2}))
differs slightly from the value  $27.22(11)$ obtained in the work
\cite{vasilyev01}.
Note also that the value (\ref{eq:beta2}) coincides
with the value presented in \cite{dzuba97} which was obtained with slightly
different E1 amplitudes. This is because of an accidental cancellation of
the changes to different terms.

Using the conversion $|e|/a_{B}^{2}=5.1422\times 10^{12}{\rm mV}/{\rm cm}$,
we therefore obtain for the weak charge of the Cs nucleus
with $\beta =26.957$:
\begin{equation}
\label{eq:q1}
Q_{W}=-72.09(29)_{\rm exp}(36)_{\rm theor} \ ,
\end{equation}
or with $\beta =27.15$
\begin{equation}
\label{eq:q2}
Q_{W}=-72.60(39)_{\rm exp}(36)_{\rm theor} \ ,
\end{equation}
where the experimental error is obtained by adding in quadrature the
error for $\beta$ and the error for ${\rm Im}(E_{PNC})/\beta$.

If we take an average value for $\beta$
\begin{equation}
	\label{eq:beta3}
	\beta =26.99(5) a_{B}^{3}
\end{equation}
then
\begin{equation}
\label{eq:q3}
Q_{W}=-72.18(29)_{\rm exp}(36)_{\rm theor} \ .
\end{equation}

These results (Eqs.~(\ref{eq:q1},\ref{eq:q2},\ref{eq:q3})) deviate by
$2.2\sigma$, $0.9\sigma$ and $2.0\sigma$, respectively, from
the standard model value $Q_{W}= -73.09(3)$ \cite{groom00}
(note that the standard deviations $\sigma$ are different
in each case).

\acknowledgments

We are grateful to A. Milstein, O. Sushkov, and M. Kuchiev for
useful discussions.
This work was supported by the Australian Research Council.


\begin{table}
\caption{Contributions to the $6s-7s$ $E_{PNC}$ amplitude
for Cs in units $10^{-11}iea_{B}(-Q_{W}/N)$.
($\hat{\Sigma}$ corresponds to the (unfitted) ``dressed''
self-energy operator.) }
\label{tab:pnci}
\begin{tabular}{ld}
TDHF & 0.8898 \\
$\langle \psi _{7s}|\hat \Sigma_s(\epsilon_{7s})|\delta X_{6s}\rangle$
& 0.0773 \\
$\langle \delta\psi _{7s}|\hat \Sigma_p(\epsilon_{7s})|X_{6s}\rangle$
& 0.1799 \\
$\langle \delta Y_{7s}|\hat \Sigma_s(\epsilon_{6s})|\psi_{6s}\rangle$
& -0.0810 \\
$\langle Y_{7s}|\hat \Sigma_p(\epsilon_{6s})|\delta \psi_{6s}\rangle$
& -0.1369 \\
Nonlinear in $\hat{\Sigma}$ correction & -0.0214 \\
Weak correlation potential & 0.0038 \\
Structural radiation & 0.0029 \\
Normalization & -0.0066 \\
 & \\
Subtotal & 0.9078 \\
 & \\
Breit & -0.0055 \\
Neutron distribution & -0.0018 \\
Radiative corrections & 0.0036 \\
 & \\
Total & 0.9041 \\
\end{tabular}
\end{table}
\begin{table}
\caption{Radiative corrections to RHF removal energies; units cm$^{-1}$.
(See also Table~\ref{tab:energies}.) }
\label{tab:radenergies}
\begin{tabular}{dddd}
$6s$ & $7s$ & $6p_{1/2}$ & $7p_{1/2}$ \\
\hline

-18.4 & -5.0 & 0.88 & 0.31 \\
\end{tabular}
\end{table}
\begin{table}
\caption{Removal energies for Cs in units cm$^{-1}$.}
\label{tab:energies}
\begin{tabular}{lllll}
State & RHF & $\hat{\Sigma} ^{(2)}$ & $\hat{\Sigma}$ &
Experiment \tablenotemark[1]\\
\hline
$6s$ & 27954 & 32415 & 31492 & 31407 \\
$7s$ & 12112 & 13070 & 12893 & 12871 \\
$6p_{1/2}$ & 18790 & 20539 & 20280 & 20228 \\
$7p_{1/2}$ & 9223 & 9731 & 9663 & 9641 \\
\end{tabular}
\tablenotetext[1]{Taken from \cite{moore}.}
\end{table}
\begin{table}
\caption{Radial integrals of E1 transition amplitudes for Cs in
different approximations. The experimental values are listed in
the last column. (a.u.)}
\label{tab:e1i}
\begin{tabular}{ldddddd}
Transition & RHF& TDHF & $\hat{\Sigma} ^{(2)}$ &
$\hat{\Sigma}$ & $\hat{\Sigma}$ & Experiment \\
       &      &              & with fitting &           & with fitting &
 \\
\hline
$6s-6p$ & 6.464 & 6.093 & 5.499 & 5.497 & 5.509 &
5.497(8) \tablenotemark[1] \\
$7s-6p$ & 5.405 & 5.450 & 5.198 & 5.190 & 5.204 &
5.185(27) \tablenotemark[2]\\
$7s-7p$ & 13.483 & 13.376 & 12.602 & 12.601 & 12.612 &
12.625(18) \tablenotemark[3] \\
\end{tabular}
\tablenotetext[1]{Ref. \cite{rafac99}.}
\tablenotetext[2]{Ref. \cite{bouchiat84}.}
\tablenotetext[3]{Ref. \cite{bennett99}.}
\end{table}
\begin{table}
\caption{Percentage deviation from experiment of calculated
E1 transition amplitudes in different approximations.}
\label{tab:e1ii}
\begin{tabular}{lddd}
Transition &
\multicolumn{3}{c}{Percentage deviation} \\
 & $\hat{\Sigma}^{(2)}$ & $\hat{\Sigma}$ & $\hat{\Sigma}$ \\
 & with fitting    &          & with fitting \\
\hline
$6s-6p$ & 0.04 &  0.0 &  0.2 \\
$7s-6p$ &  0.3 &  0.1 &  0.4 \\
$7s-7p$ & -0.2 & -0.2 & -0.1 \\
\end{tabular}
\end{table}
\begin{table}
\caption{Calculations of the hyperfine structure of Cs
in different approximations. In the last column the experimental
values are listed. Units: MHz.}
\label{tab:hfsi}
\begin{tabular}{ldddddd}
State & RHF & TDHF & $\hat{\Sigma} ^{(2)}$ &
$\hat{\Sigma}$ & $\hat{\Sigma}$ & Experiment \\
       &    &     & with fitting &      & with fitting & \\
\hline
$6s$ & 1425.0 & 1717.5 & 2306.9 & 2315.0 & 2300.3 &
2298.2 \tablenotemark[1] \\
$7s$ & 391.6 & 471.1 & 544.4 & 545.3 & 543.8 &
545.90(9) \tablenotemark[2] \\
$6p_{1/2}$ & 160.9 & 200.3 & 291.5 & 293.6 & 290.5 &
291.89(8) \tablenotemark[3] \\
$7p_{1/2}$ & 57.6 & 71.2 & 94.3 & 94.8 & 94.1 &
94.35 \tablenotemark[1] \\
\end{tabular}
\tablenotetext[1]{Ref. \cite{cshfs}.}
\tablenotetext[2]{Ref. \cite{gilbert83}.}
\tablenotetext[3]{Ref. \cite{rafac97}.}
\end{table}
\begin{table}
\caption{Percentage deviation from experiment of calculated
hyperfine structure constants in different approximations.}
\label{tab:hfsii}
\begin{tabular}{ldddd}
State &
\multicolumn{3}{c}{Percentage deviation} \\
 & $\hat{\Sigma}^{(2)}$ & $\hat{\Sigma}$ & $\hat{\Sigma}$ \\
 & with fitting    &          & with fitting \\
\hline
$6s$       &  0.4 &  0.7 &  0.09 \\
$7s$       & -0.3 & -0.1 & -0.4 \\
$6p_{1/2}$ & -0.1 &  0.6 & -0.5 \\
$7p_{1/2}$ & -0.05 & 0.5 & -0.3 \\
\end{tabular}
\end{table}
\begin{table}
\caption{Percentage deviation from experiment of calculated
$\sqrt{{\rm hfs}(s){\rm hfs}(p)}$ (we will denote this by $s-p$ in
the tables) in different approximations.}
\label{tab:hfsiii}
\begin{tabular}{lddd}
$\sqrt{{\rm hfs}(s){\rm hfs}(p)}$&\multicolumn{3}{c}{Percentage deviation}\\
 & $\hat{\Sigma} ^{(2)}$ & $\hat{\Sigma}$ & $\hat{\Sigma}$ \\
 & with fitting    &          & with fitting \\
\hline
$6s-6p$ &  0.1 &  0.7 & -0.2 \\
$6s-7p$ &  0.2 &  0.6 & -0.09 \\
$7s-6p$ & -0.2 &  0.2 & -0.4 \\
\end{tabular}
\end{table}
\begin{table}
\caption{Values for $E_{PNC}$ in different approximations;
units $10^{-11}iea_{B}(-Q_{W}/N)$.}
\label{tab:pncii}
\begin{tabular}{cddd}
 & $\hat{\Sigma}^{(2)}$ with fitting & $\hat{\Sigma}$ &
$\hat{\Sigma}$ with fitting \\
\hline
$E_{PNC}$ & 0.901 & 0.904 & 0.903 \\
\end{tabular}
\end{table}

\center
\widetext
\input psfig
\psfull

\end{document}